# Time-domain separation of optical properties from structural transitions in resonantly bonded materials

*Lutz Waldecker[1] \*, Timothy A. Miller[2]\*, Miquel Rudé[2], Roman Bertoni[1], Johann Osmond[2], Valerio Pruneri[2,3], Robert E. Simpson[4], Ralph Ernstorfer[1]†, Simon Wall[2]‡*

*1 Fritz-Haber-Institut der Max-Planck-Gesellschaft, Faradayweg 4-6, D-14195 Berlin, Germany*

*2 ICFO—Institut de Ciències Fotòniques, Mediterranean Technology Park, 08860, Castelldefels, Barcelona, Spain*

*3 ICREA—Institució Catalana de Recerca i Estudi Avançats, 08015 Barcelona, Spain*

*4 SUTD— Singapore University of Technology & Design, 8 Somapah Road, 487372, Singapore*

**The extreme electro-optical contrast between crystalline and amorphous states in phase change materials is routinely exploited in optical data storage[1] and future applications include universal memories[2], flexible displays[3], reconfigurable optical circuits[4,5], and logic devices[6]. Optical contrast is believed to arise due to a change in crystallinity. Here we show that the connection between optical properties and structure can be broken. Using a unique combination of single-shot femtosecond electron diffraction and optical spectroscopy, we simultaneously follow the lattice dynamics and dielectric function in the phase change material Ge$_2$Sb$_2$Te$_5$ during an irreversible state transformation. The dielectric function changes by 30% within 100 femtoseconds due to a rapid depletion of electrons from resonantly-bonded states. This occurs without perturbing the crystallinity of the lattice, which heats with a 2 ps time constant. The optical changes are an order-of-magnitude larger than those achievable with silicon and present new routes to manipulate light on an ultrafast timescale without structural changes.**

Ge$_2$Sb$_2$Te$_5$ (GST) is the prototypical phase change material and exhibits the fastest crystallization[7] and amorphization rates[6] measured to date. The crystalline state is electrically conductive and optically opaque whereas the amorphous state lacks long-range order and has a lower electrical conductivity and optical absorption. Understanding the speed of the structural transformation and its relation to the optical properties continues to be a significant research topic for improving the design of materials for future devices.

The conventional pathway for amorphization is to heat the crystalline state above the melting point (T$_m$ = 615 °C)[8] to disorder the lattice and then rapidly quench the system to freeze the disorder[9]. However, in order to explain the speed of amorphization in GST, alternative mechanisms have been proposed. It has been suggested that the Ge ions' coordination changes through an 'umbrella flip' transition, which modifies both the structure and optical properties[10]. As only a subset of bonds would change, this type of mechanism has the potential to be very fast and non-thermal. However, the thermal stability of the umbrella flip model has been questioned,[11] and later work has shown that coordination changes alone are insufficient to explain the observed changes in the optical properties[12,13].

Alternatively, the crystalline-amorphous optical contrast has been explained in terms of resonant bonds in the crystalline state[12,13]. Resonant bonds form in specific crystalline systems in which the electronic orbitals of half-filled p-type bands are aligned over next-nearest neighbours. These extended delocalized states give rise to a large dipole matrix-element enhancement of the optical properties which is manifested in the large value of the real part of the dielectric function in the low energy (<2 eV) region[14]. This enhancement is lost when there is angular disorder between the extended p-states, changing the optical properties drastically despite the local (nearest-neighbour) bonding remaining relatively unperturbed, as is the case for amorphous GST[13,15].

---

\* These authors made equal contribution
† ernstorfer@fhi-berlin.mpg.de
‡ simon.wall@icfo.es

The sensitivity of resonant bonds to bond alignment has also generated ideas for non-thermal transformation routes. Simulations have shown that distortions to a subset of bonds in the crystalline state of phase change materials can trigger collapse to the amorphous state[16]. As thermal heating affects all bonds, selective excitation of specific bonds could be a more efficient route to generate the amorphous phase, and it has been suggested that this may occur during the photo-driven phase transformation.

Here, we provide the first direct measurements of structural and optical dielectric properties during the photo-triggered amorphization of GST using a combination of time-resolved femtosecond optical spectroscopy and electron diffraction in the single-shot regime. This combination allows us to unambiguously disentangle the electronic and structural contributions to the optical properties. In the first few picoseconds, we observe a separation of the optical properties from the structural state. Changes in the dielectric function, very similar to the equilibrium changes between crystalline and amorphous states, are observed immediately after excitation, consistent with a loss of resonant bonding. However, on this time scale the lattice is still cold and long-range order is still present. Permanent switching to the amorphous state is a slower process, dictated by the thermal response of the lattice that only occurs when the induced temperature jump is sufficient to melt the film.

Figure 1a shows the permanent change induced in GST by single 35-fs pulses at 800-nm central wavelength and 30 nm bandwidth. We find that an incident fluence greater than $F_{TH}$ = 14 mJ cm$^{-2}$ leads to amorphization of the 30 nm thick crystalline thin-film samples, and fluences above 32 mJ cm$^{-2}$ result in ablation. Amorphization was confirmed by a Raman analysis of the pumped region, which is shown in the supplementary information together with the sample preparation and characterization techniques.

The dynamics of the crystalline state for fluences below $F_{TH}$ are shown in Figure 1b. These measurements were made using conventional pump-probe techniques at low repetition rates (<100 Hz) to avoid cumulative heating in the sample. Although the single-shot transformation threshold was 14 mJ cm$^{-2}$, the maximum fluence in the reversible regime was limited to approximately 7 mJ cm$^{-2}$, as a gradual transformation of the sample over several minutes was observed for higher fluences. We simultaneously measure the time-dependent changes in the transmitted and reflected light at 800 nm, which we use to calculate the time-dependent dielectric function shown in Figure 1b. Electron diffraction patterns (Figure 1c) were measured in transmission with 92 keV femtosecond electron pulses[17] and the diffraction peaks were fitted to obtain their evolution as a function of time delay (Figure 1d). All measurements were performed at near-normal incidence, and further details can be found in the methods and supplementary information.

For below threshold excitation we observe 3 phenomena: *1.* Photo-excitation triggers a pronounced and prompt decrease in both the real ($\varepsilon_1$) and imaginary ($\varepsilon_2$) parts of the dielectric function followed by an exponential recovery, whereas the lattice shows a slow decrease in diffraction intensity. *2.* An oscillatory response modulates $\varepsilon_2$. *3.* The recovery time of the optical signal *increases* with *increasing* pump fluence, whereas the lattice dynamics are *independent* of fluence (Figure 1e).

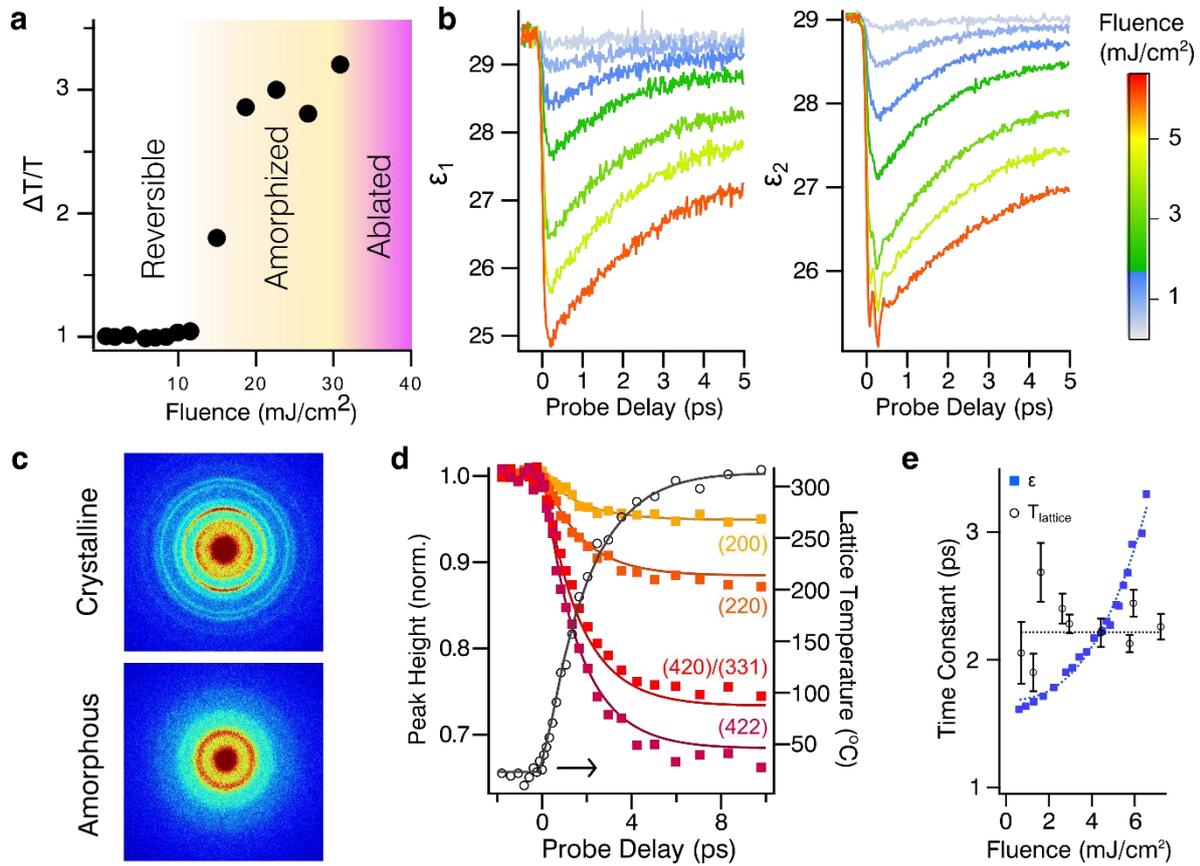

**Figure 1 | Below-threshold dynamics in crystalline GST a.** Permanent change in sample transmission measured minutes after irradiation by single femtosecond laser pulses. The sample amorphizes with a single pulse for F> $F_{TH}$= 14 mJ cm$^{-2}$ while ablation occurs for F > 32 mJ cm$^{-2}$. **b.** Dynamics of the real and imaginary parts of dielectric function at 1.5 eV show a prompt decrease and exponential recovery after photoexcitation. **c.** Static diffraction patterns of crystalline and amorphous states recorded with femtosecond electron bunches at 92 keV. **d.** Evolution of several diffraction peaks after excitation with 5.8 mJ cm$^{-2}$ and the extracted temperature change. The temperature dynamics are fit with a single 2.2 ps exponential rise. **e.** The time constant from the diffraction and the optical data for various fluences showing the recovery time of ε is sensitive to the fluence, whereas the heating dynamics are independent of fluence.

Optically exciting GST at 800 nm generates free carriers by interband transitions. The enhanced free-carrier density increases the plasma frequency, which *decreases* $\varepsilon_1$ and *increases* $\varepsilon_2$. When observed at 800 nm, these effects are usually small[18]. However, the optical response of crystalline GST shows a large decrease in both $\varepsilon_1$ *and* $\varepsilon_2$ and is thus not described by the response of free carriers. Instead, we attribute this effect to a photo-bleaching of electrons from resonantly bonded states, as the primary effect resulting from the loss of resonant bonding is a large decrease in ε[12]. The initial decrease of $\varepsilon_1$ and $\varepsilon_2$ scales linearly with pump intensity; however, the recovery time, which is the same for $\varepsilon_1$ and $\varepsilon_2$, increases non-linearly as shown in Figure 1e. Photoexcitation also generates oscillations which only perturb $\varepsilon_2$, corresponding to coherent Raman-active vibrations (see the supplementary information for a comparison of the spectra)[19]. As the Raman-active modes do not significantly perturb $\varepsilon_1$, we conclude that they do not particularly perturb the resonant bonds.

We now consider the evolution of the diffraction peaks following laser excitation shown in Figure 1d. The temporal behaviour can be explained by an increase in mean-square displacement of the atoms around their equilibrium position. If we assume that the lattice vibrations are thermal at all times, the system is describable by a time-dependent temperature, which can be obtained through an analysis of the Debye-Waller B-factor[20]. The extracted temperature evolution is also shown in Figure 1d and can be fitted using a single exponential rise with a time constant of 2.2 ps (see methods for details). The good agreement between the fitted diffraction peaks based on a thermal lattice and the measured data suggests that the lattice is, to good approximation, thermal.

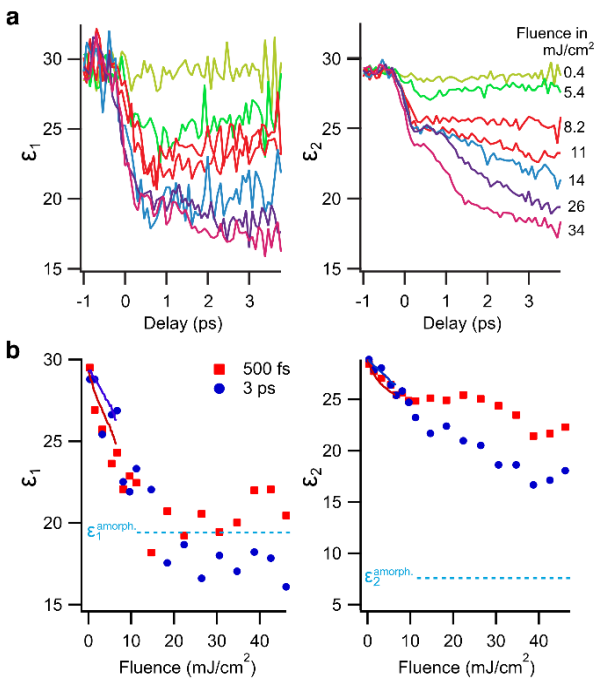

**Figure 2 | Dynamics of the dielectric function during amorphization. a.** Time dependence of ε, labelled by the incident fluence in mJ/cm². As $F_{TH}$ is approached, $\varepsilon_1$ saturates, but $\varepsilon_2$ shows a further slow dynamic. Increasing the fluence of the pump pulse above the ablation threshold (F> 32 mJ/cm²) does not cause a significant change in either dynamic. **b.** The power dependence highlights the saturation of $\varepsilon_1$ above $F_{TH}$ at the value of the dielectric function in the amorphous phase (horizontal dashed line). This occurs within the time resolution and shows little subsequent evolution. $\varepsilon_2$ saturates at short delays at a value far from the equilibrium amorphous value, but continues to evolve on longer timescales. Solid lines show low fluence data from Fig 1.

Interestingly, unlike the optical signal, the lattice-heating time constant is independent of fluence (Figure 1e). This difference indicates that the recovery of the dielectric function cannot be simply attributed to electrons thermalizing with the lattice. Instead, it can be understood in terms of resonant bonding. As the lattice heats, ions fluctuate further from their equilibrium positions, disrupting the p-orbital alignment. Since lattice disorder makes bond alignment over an extended number of lattice sites unlikely, reformation of resonant bonding becomes less probable and the recovery time increases as the lattice heats.

When the pump power is increased above $F_{TH}$, single pump pulses induce irreversible changes in the sample, and conventional pump-probe techniques are not suitable to study the dynamics. Instead the optical measurements harness a chirped probe pulse and spectral encoding to map the temporal response onto the frequency domain, allowing us to measure the initial 3-ps dynamics with 100 fs time resolution using only a single laser pulse[21]. Electron diffraction images were taken with single electron bunches, and the dynamics were measured by moving the sample to a fresh spot for each time point. Technical details are given in the supplementary information.

Figure 2 shows the dynamics of the dielectric function for a range of fluences that cover amorphization and even ablation regimes. The amplitude of the initial drop of $\varepsilon_1$ increases with increasing excitation fluence until the threshold fluence. At this point, $\varepsilon_1$ has decreased by 30% within 100 fs and does not respond to further increases in pump fluence nor evolve in time. Remarkably, the saturated value reached on the femtosecond timescale is the same as that of the amorphous phase. As resonant bonds are not present in the amorphous phase, the appearance of properties similar to the amorphous state indicates that the resonant-bonding contribution to the dielectric function has been completely suppressed within 100 fs. $\varepsilon_2$ also shows a similar saturation as a function of fluence at short times, but at a value that is still far from that of the amorphous phase. Unlike $\varepsilon_1$, however, it continues to evolve in time, and large changes occur on a slower timescale.

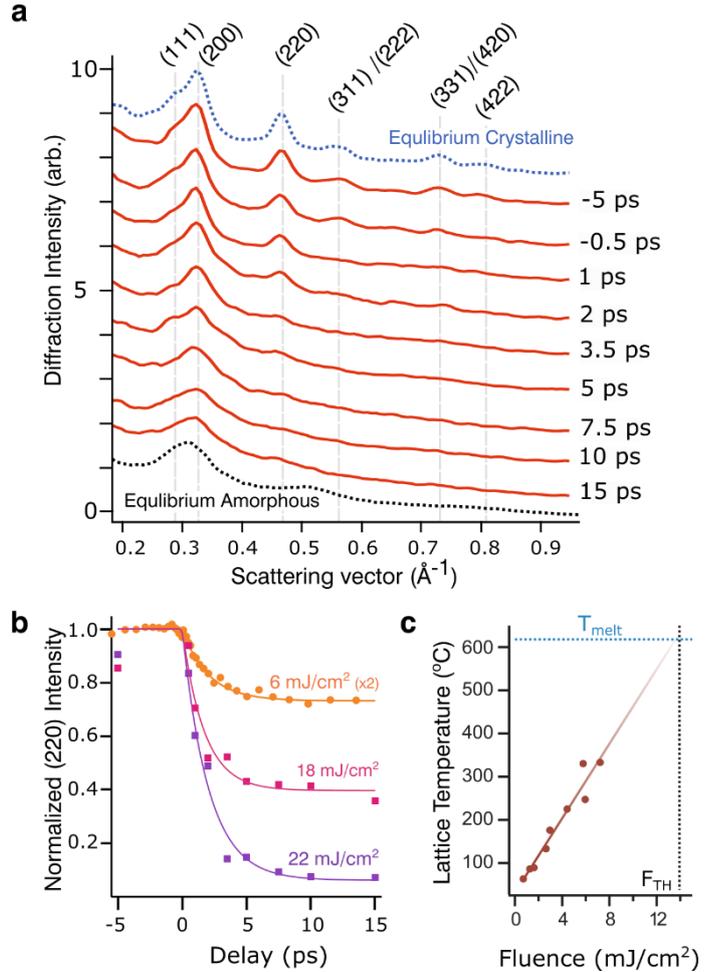

**Figure 3 | Structural dynamics during amorphization. a.** Raw radially integrated diffraction patterns measured in the single-shot regime for a pump fluence of 22 mJ/cm² at several time delays. The diffraction orders are labelled, with multiple labels indicating that diffraction orders overlap. Contrast between the (200) and (220) peaks is lost by approximately 5 ps. **b.** The time evolution of the (220) diffraction peak. The same time profile is obtained using the below-threshold data of the lattice heating with a time constant of 2.2 ps to fit the above-threshold response by scaling the amplitude. This demonstrates that the loss of long-range order occurs by thermal melting. **c.** Fluence dependence of the calculated temperature after laser excitation from the below threshold data. The straight line is a linear fit to the low fluence data.

The ultrafast saturation in $\varepsilon_1$ to the amorphous phase value suggests that the resonantly bonded states have been completely suppressed on a sub-picosecond timescale as would be expected for a non-thermal phase transition. Femtosecond electron diffraction, however, reveals that the lattice responds on a slower timescale. Figure 3a shows that the crystalline structure persists for several picoseconds after photoexcitation, and the complete loss of long-range order is observed after approximately 5-10 ps. Figure 3b shows the temporal evolution of the (220) peak for 3 fluences. The temporal evolution above threshold can be fit by scaling the amplitude of the intensity change and keeping the timescales constant, indicating that the above-threshold lattice dynamics are described by the same thermal process. Figure 3c shows that the final lattice temperature achieved for below threshold excitation increases linearly with excitation fluence. Although this linearity is not expected to continue when the state changes, extrapolating up to the threshold reveals a good agreement between the melting temperature and the transformation fluence.

The unique combination of measurements presented here allows us to obtain a microscopic picture of the initial steps in the photo-induced amorphization process in phase change materials, which is schematically depicted in Figure 4. Femtosecond optical excitation directly removes electrons from resonantly bonded states as evidenced by the immediate decrease in the dielectric function and the saturation of $\varepsilon_1$ at the amorphous-state value. This non-thermal femtosecond change in the optical properties does not coincide with a change in crystallinity and represents a previously unobserved non-equilibrium state of GST.

This non-equilibrium state is lost during lattice heating. Heating occurs with a 2.2 ps exponential time constant which is dictated by the rate of energy transfer from electrons to vibrations of the covalently-bonded backbone of the crystalline lattice[16]. When photoexcited with 800 nm light to the level needed to completely suppress resonant bonding, there is also sufficient energy deposited to melt the crystal upon thermalization. Once melted, the final state obtained depends on how quickly the heat is extracted from the liquid. We find no evidence for any non-thermal lattice dynamics, in contrast to observations in other materials[22–25] and thus photoexcitation at 800 nm is unable to affect the lattice non-thermally as described in reference 16. Such dynamics may be observable at higher excitation fluences in the ablation regime, which is of limited interest for practical reasons. Our combination of optical and structural probing allows us to deduce that structural amorphization is much slower than the 130-200 fs suggested from optical transmission data alone[26]. Phase transformation through the molten state is also consistent with previous time-resolved measurements on photo-induced crystallization in another phase change material, GeSb, in which crystallization occurred after the amorphous state was melted[27].

The observation that resonant bonds can be controlled rapidly and non-thermally, apart from the crystalline lattice, represents a new understanding of the phase transformation in GST. Switching the optical properties of GST is no longer dictated by the absolute atomic arrangement, but rather only stabilized by it. Without the complete depopulation of the resonantly bonded state, we achieve reversible modulations of the dielectric function of up to 13%, over an order of magnitude larger than observed in silicon photo-switches[28]. By extracting the energy used to depopulate the resonant bonds before the lattice heats above the melting temperature, as suggested in Figure 4, an ultrashort-lived state with 30% change in the dielectric function may be achieved. Energy extraction could be realized in nanostructured devices by rapid transfer of the photo-excited carriers, both electrons and holes, into a metal or semimetal, for instance in layered heterostructures composed of GST and graphene or few-layer graphite. Additionally, efficient vibrational coupling of nanoscale GST to the surrounding will limit the maximum lattice temperature, and resonant bonds could re-establish and recover the crystalline state optical properties on the few-ps time scale. As the structural transition limits the lifetime of GST-based devices, achieving optical contrast without the structural change is a major advantage and will improve cyclability. We expect that the transient optical changes will be even larger at telecommunication wavelengths, as the infrared spectral region has a greater sensitivity to changes in resonant bonding; thus, the ability to harness the ultrafast optical contrast of phase change materials without structural transition suggests a new avenue to high speed optical modulators[29] for communications and computations.

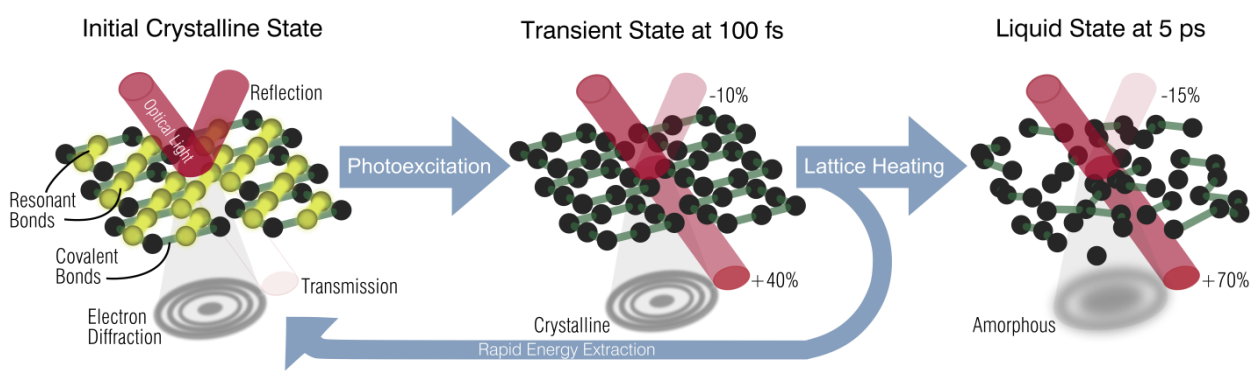

**Figure 4 | Schematic of the ultrafast transformation pathway.** Crystalline GST exists with resonant bonds (yellow, subset shown for clarity) which dictate the optical properties. Ultrafast photo-excitation forms a new transient state by removing the resonant bonds and changing the optical properties before ionic motion has occurred. Transfer of the electron energy to the covalent backbone causes lattice heating, which thermally melts the long-range order after several picoseconds. Subsequent thermalization determines the final material state. Rapid energy extraction during the transient state could prevent thermal melting and enable the restoration of the resonantly bonded state quickly.

# Methods

### Sample Preparation

For optical measurements 30 nm GST films were grown on fused silica substrates by RF co-sputtering from two stoichiometric targets of GeTe and $Sb_2Te_3$ in an Ar atmosphere. The as-deposited film is amorphous and its stoichiometry is confirmed to be close to $Ge_2Sb_2Te_5$ by EDX measurements (accuracy of 5 %). 10 nm $Si_3N_4$ capping layer films were grown by reactively DC sputtering from a Si target for 10 min using a mixture of Ar and $N_2$ as a process gas. The capping layer is used to prevent oxidation of the GST layer.

Crystallization of the GST was achieved by annealing the sample on a hot plate at 200 °C for 1 h using a heating rate of 10 °C/min, and the final state was characterized by Raman and optical spectra.

### Determination of the dielectric function

The dielectric function was measured using broadband ellipsometry. The values obtained were in excellent agreement at 800 nm to those found using a transfer matrix method to invert reflection (R) and transmission (T) data measured at normal incidence on the same films.

Time dependent measurements measure ΔR/R and ΔT/T. We use the static R and T values to convert the transients into the absolute value of R and T as a function of time. These values were then inverted using the transfer matrix method to recover the complex dielectric function. We assume coherent reflections from the GST and $Si_3N_4$ layers ($n_{Si_3N_4}$ = 1.99622) and incoherent reflections from the backside of the $SiO_2$ substrate ($n_{SiO_2}$ = 1.45332). Raw reflection and transmission data and further details can be found in the supplementary information.

### Single-Shot Optical Technique

Single-shot measurements were performed using a spectral-encoding technique[21]. A 35 fs, 800 nm pulse passes twice through block of glass, stretching the pulse to approximately 12 ps with a linear chirp. The pulse is incident on the sample at near-normal incidence and, together with a reference signal taken before the sample, the transmitted and reflected signals are focused into separate fibers and are individually imaged onto a CCD camera in an imaging spectrometer.

The linear chirp results in each color in the probe interacting with the sample at a different time. If we assume that the resulting ΔR/R and ΔT/T due to the time dependent dielectric function is independent of the probe frequency within our laser spectrum, we can extract the time-dependent signal. The details of the temporal calibration are found in the supplement.

### Femtosecond Electron diffraction

Diffraction experiments use a pump-probe scheme with an 800 nm 35 fs pump-pulse and a short electron bunch as a probe of the structural state. We create short electron pulses by photo-emission of electrons from a photocathode and acceleration to a kinetic energy of 92 keV in a static electric field. The electron bunch then diffracts off the sample that is placed shortly behind the anode, and a two-dimensional diffraction image is recorded with an electron camera. For the employed bunch charge, the electron pulse duration on the sample is estimated to be 150 fs from measurements made on Al foils.

Due to the high scattering cross-section of electrons with matter, the diffraction experiment requires thin, freestanding films of the sample under study. To ensure comparability to the optical experiments, we use a sandwich-geometry consisting of a 30 nm thick film of GST with a 10 nm thick capping layer of $Si_3N_4$ on the front side and replace the glass substrate with yet another 10 nm thick layer of $Si_3N_4$. The films are transferred to a large area (20 x 20 mm$^2$) grid with 190 x 190 μm² holes etched from a silicon wafer. The sample can be translated in three directions in the vacuum chamber and its tip and tilt can be controlled by two additional motors to ensure parallel movement to the stages. In the single-shot experiment, a fresh piece of sample can be supplied for each laser shot without changing the temporal overlap of pump and probe.

Electron diffraction experiments with reversible excitation have been performed at different repetition rates to avoid temperature offsets at negative delay times due to accumulated heating by multiple pump pulses. By comparison of diffraction images at negative pump-probe delays to measurements without pump laser, the repetition rate was lowered until no offset could be observed.

ELECTRON DIFFRACTION DATA ANALYSIS AND LATTICE TEMPERATURE CALCULATION

We record two-dimensional diffraction images with a phosphor screen fiber-coupled to a CMOS chip (TVIPS TemCam-F416). The appearance of weak texture in the Debye-Scherrer rings has been observed in some samples (see figure 1d), but has not shown to influence the structural dynamics. The recorded images are thus integrated angularly in both single-shot and reversible measurements to obtain radial averages. From these, the relative peak heights can be determined by fitting and subtracting a background (Lorentzian + fourth order polynomial) and fitting pseudo-Voigt line profiles to the peaks for each time delay.

A Debye-Waller type analysis of the diffraction data was used to calculate a lattice temperature from the relative diffraction peak heights. This analysis relies on the assumption that the phonons are close to a thermal distribution at all delay times and that phonon modes do not shift through photoexcitation, which we find to be valid.

The relative intensity of a peak with scattering vector S at a temperature T with respect to the initial reference temperature, $T_0$, is given by:

$$I_{rel,S}[T(t)] = \exp\left(S^2\big(B(T_0) - B(T(t))\big)\right)$$

with B(T) being the Debye-Waller B factor for the given temperature which is given in ref [20]. Equivalently, the relative intensity of a Bragg peak with scattering vector S of a sample in a thermal state can be written in a form that only depends on temperature:

$$-\frac{\ln\left(I_{rel,S}(t)\right)}{S^2} = B\big(T(t)\big) - B(T_0)$$

Plotted in this form, the relative intensities of all Bragg peaks lie on top of each other. This indicates that the phonon distribution is isotropic and no strong coupling to specific modes is present. We therefore calculate a lattice temperature T(t) from our data, which is obtained by averaging the peak intensities in this form and numerically solving the equation. This temperature T(t) is used to recalculate the expected evolution of the individual peaks, and it is this calculated peak height as a function of temperature that is plotted with solid lines in Figure 1d.

Finally T(t) is fitted by the following fit function:

$$T(t) = T_0 + \Delta T_1 \left(1 - \exp\left(-\frac{t}{\tau_1}\right)\right) - \Delta T_2 \left(1 - \exp\left(-\frac{t}{\tau_2}\right)\right), t > 0$$

where $T_0$ is initial lattice temperature, $\Delta T_1$ is the temperature increase resulting from the thermalization of the electrons and the lattice and $\tau_1$ is the heating rate, plotted in Figure 1e. $\Delta T_2$ is the amplitude of cooling, and $\tau_2$ is the cooling rate, due to heat diffusion into the $Si_3N_4$ capping layers. $\tau_2$ was found from long time scans of the diffraction signal and was found to be 100 ps. Heat diffusion resulted in a cooling of the film by 20-50% of the initial heating at 100 ps. The values of $\Delta T_1$ found in the Debye-Waller analysis are in good agreement with expected temperature rises using values of the static heat capacity of 0.25 Jg$^{-1}$K$^{-1}$ and the absorbed energy densities of the GST films[30].

ACKNOWLEDGMENTS

TAM acknowledges funding through the Marie Curie COFUND project and Spanish Ministry of Economy and Competitiveness (MINECO). RB thanks the Alexander von Humboldt Foundation for financial support. VP acknowledges financial support from MINECO and the "Fondo Europeo de Desarrollo Regional" (FEDER) through grant TEC2013-46168-R. RE acknowledges fruitful discussions with M. Wuttig and funding from the Max Planck Society. SW acknowledges financial support from Ramon y Cajal program RYC-2013-14838 and Marie Curie Career Integration Grant PCIG12-GA-2013-618487.

## Author Contributions

SW, LW and RE initiated the project. TAM and SW performed the multi-shot optical measurements. SW, LW and TAM performed the single shot optical measurements. LW and RB performed the time resolved diffraction measurements. MR fabricated samples, which were characterized by MR, TAM and JO. All authors provided input to the interpretation of the data and writing the manuscript.

## Supplemental Information

The supplemental information contains the measured reflection and transmission data, the conversion procedure to the dielectric function and the fitting, as well as information about the single-shot timing calibration and additional details about sample characterization.

## Competing Financial Interests

The authors declare no competing financial interests.

# TIME-DOMAIN SEPARATION OF OPTICAL PROPERTIES FROM STRUCTURAL TRANSITIONS IN RESONANTLY BONDED MATERIALS

*Lutz Waldecker, Timothy A. Miller, Miquel Rudé, Roman Bertoni, Johann Osmond, Valerio Pruneri, Robert Simpson, Ralph Ernstorfer, Simon Wall*

## ELLIPSOMETRY AND RAMAN ANALYSIS OF PHASE TRANSFORMATION

Figure S1a shows the measured dielectric function for three thermally prepared states. The amorphous state achieved after annealing to 75 °C is similar to the one achieved after quenching from the crystalline state. The values for the dielectric function reported here are different to those by Shportko et al[1]; however, we find a much better agreement with the dielectric functions found in references 2 and 3. This variation is likely due to the metastability of both crystalline and amorphous phases of GST. Small variations in composition and preparation can alter the state of the material, and substrates and capping layers can induce strain, resulting in slightly different optical values. However, in all cases the energy dependence of the dielectric function follows the same trend. Figure S1b shows the Raman spectra for the different states, which are in excellent agreement with other publications[4–7]. The spectrum obtained after pumping the crystalline sample with a fluence just above the phase transformation threshold is in excellent agreement with the amorphous state spectrum, whereas regions excited with a higher fluence match the as-deposited phase spectra more closely. This indicates that the local bonding strongly depends on the temperature reached after excitation, even though long range order is lost in both cases.

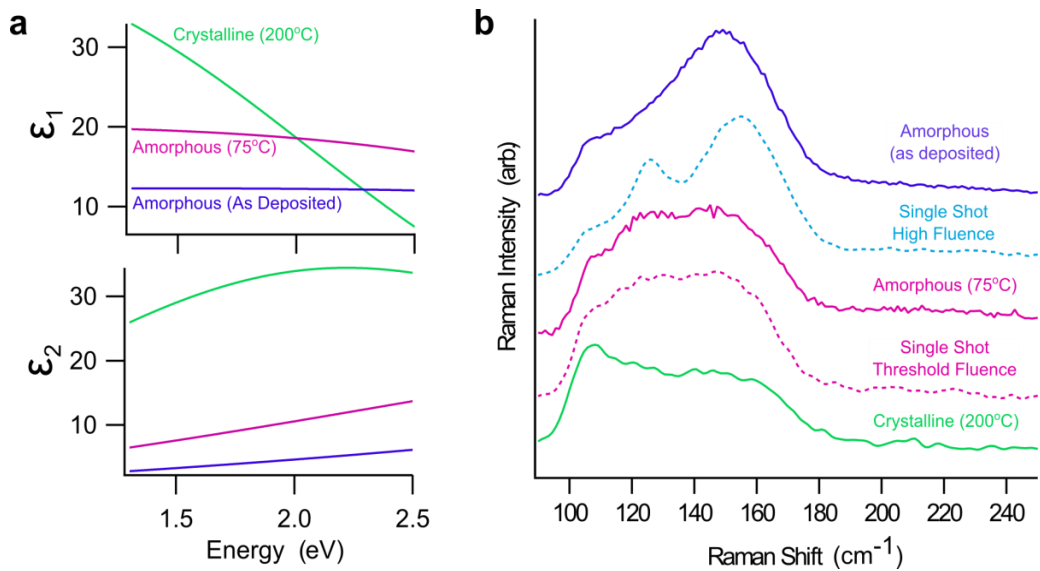

**Figure S1 | Static Characterization of GST Samples a,** Ellipsometry measurements of the dielectric function of GST in its different states prepared thermally. **b,** Raman measurements of the thermally prepared states of GST (solid lines) and the laser transformed regions (dashed lines)

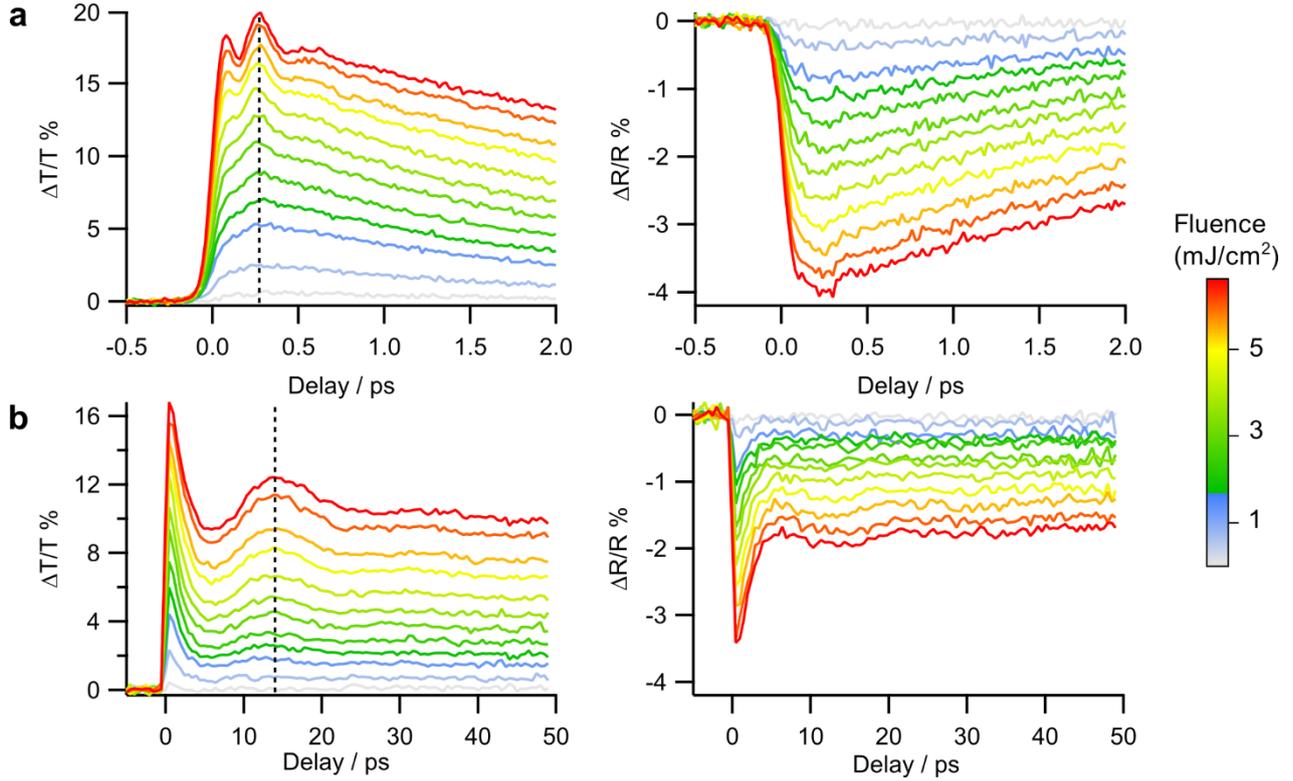

**Figure S2 | Measured Reflectivity and Transmission of GST Films Used to Calculate the Time-Dependent Dielectric Function a,** Short timescale dynamics **b,** long time dynamics. Dashed lines show the frequency independence of optical phonons (**a**) and acoustic phonons (**b**) on the pump fluence.

## BELOW-THRESHOLD DATA

The optical data in Figure 2 of the manuscript is derived from the changes in reflectivity and transmissivity of the sample measured using conventional lock-in detection. Figure S2 shows both the data on short (a) and long (b) timescales used to extract the recovery time. On short timescales we observe coherent oscillations that can be attributed to coherent Raman-active phonons (see Figure S3); these predominantly influence the transmission. The dashed line marks the end of one oscillation period and indicates that the coherent phonon frequency remains constant for all pump fluences. On longer timescales we observe a slower oscillation corresponding to the breathing motion of the film, and the frequency is set by the speed of sound of the material. As the period of oscillation for both the optical and acoustic phonons remains constant, we conclude that there is no weakening of the lattice potential in this excitation regime.

## TRANSFER MATRIX METHOD

Reflection and transmission from a layered sample can be converted into the complex dielectric function using a transfer matrix method[8]. A general transfer matrix is written as:

$$T_n = \begin{pmatrix} \cos(2\pi n \delta/\lambda) & -\frac{i}{n}\sin(2\pi n \delta/\lambda) \\ -i n \sin(2\pi n \delta/\lambda) & \cos(2\pi n \delta/\lambda) \end{pmatrix} \qquad (1)$$

where $n$ is the complex refractive index, $\delta$ is the layer thickness and $\lambda$ is the wavelength of the light. The transmission, $T$, and reflection coefficient, $R$, for the multilayer system studied here can then be obtained from the combined system $T_{SiO_2} T_{GST} T_{Si_3N_4}$ with the light entering the $Si_3N_4$ layer from air and leaving the GST layer into the $SiO_2$ substrate. A small amount of light is further reflected from the $SiO_2$-air interface. This reflection is treated incoherently, i.e. light originating from multiple reflections inside the $SiO_2$ layer does not interfere.

The time dependent dielectric function is obtained by calculating $T_{GST}$ for a range of complex refractive indices $n' = n + \Delta$, and the average is taken of the values of $n'$ that satisfy both R(t) and T(t) simultaneously. From the resulting value for $n'$ the dielectric function can be obtained. The same method is used for calculating the complex dielectric function above the amorphization threshold.

We fit the change in dielectric function with the following phenomenological model in the reversible regime:

$$\Delta\varepsilon = -Ae^{-\frac{t}{\tau}} - B\left(1 - e^{-\frac{t}{\tau}}\right), t > 0 \qquad (2)$$

where A and B are complex and $\tau$ is the recovery time of the optical signal (from Figure 1). The signal is then convolved with a Gaussian pulse of 40 fs FWHM to account for the experimental resolution. The results of the fit are shown in Figure S3 for the highest reversible fluence excitation, and an excellent fit is obtained. We attribute the A term to the dynamics of the resonantly bonded state. We can see that the dynamics of this term occur in both the real and imaginary part of the dielectric function. The coherent phonon signal, on the other hand, is only observed on the imaginary part of the dielectric function (i.e. the real part of B is small), and is in good agreement with the static Raman scattering signal.

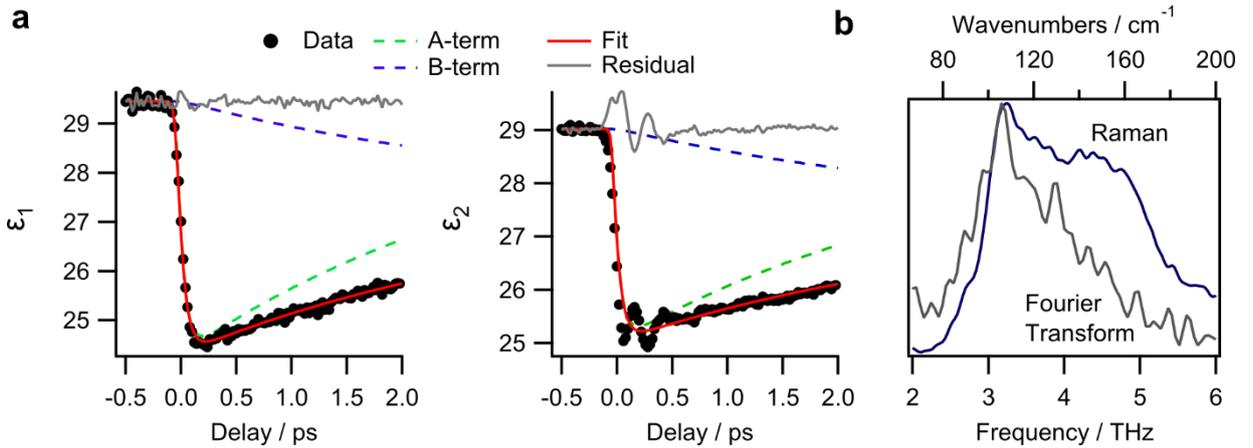

**Figure S3 | Typical Fitting results for the broadband dielectric function a,** shows the fit parameters for the two terms in Equation 2, together with the residual. **b,** The Fourier transform of the residual obtained from the imaginary part from **a**, compared to the static Raman measurement.

### SINGLE-SHOT OPTICAL TECHNIQUE

The linear chirp imparted on the probe beam, generated by passing through a thick glass block, results in a time-dependent instantaneous frequency. The pump and probe beams are timed so that roughly 1/3 of the probe beam has passed through the sample before the pump beam arrives. As a result, each color in the beam experiences a different value of the transient refractive index of the sample. If we assume that the resulting ΔR/R and ΔT/T due to the time dependent dielectric function is independent of the probe frequency within our laser spectrum, we can extract the time-dependent signal.

The frequency-to-time calibration was achieved by using a motorized delay stage to change the relative arrival time between pump and probe beams. The time zero position was found by fitting an error function to the signal at each delay as shown in Figure S4a. Figure S4b shows the extracted chirp of the pulse, which is 15.4 fs/pixel. Figure S4c shows that the ultrafast optical change of GST is independent of the probe wavelength. A limited spectral range was chosen for the dynamics due to the decrease in spectral intensity at the wings of the spectrum, resulting in more noise on the data.

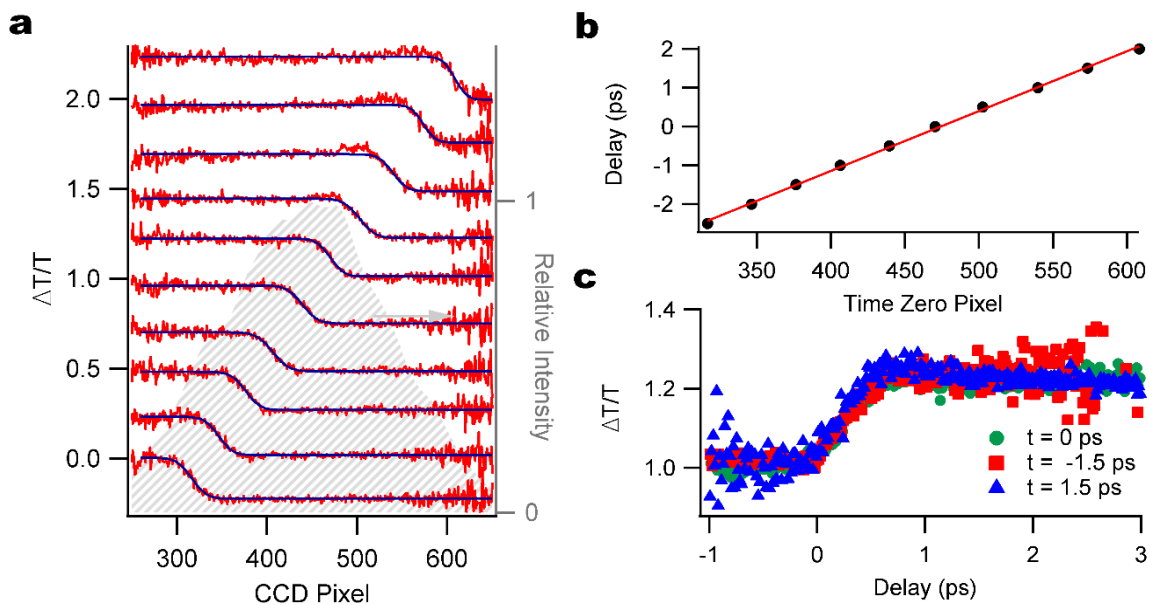

**Figure S4 | Calibration of the spectral signal into the time domain a,** The physical pump-probe delay is varied and the rising edge of the signal fit to an error function. The laser spectrum is shown in the background. **b,** The time zero point from (**a**) is plotted for different physical probe delays and shows the pulse has a linear chirp of 15.4 fs/px. **c,** The recorded dynamics are shown to be independent of wavelength within the measured spectral range. After shifting the time traces by the appropriate time delay, the exact same dynamics are observed. Note that the increased noise at negative delays for t=-1.5 ps and positive delays for 1.5ps is due to the loss of intensity at the edges of the spectrum.

To measure the irreversible dynamics we perform three measurements at each position on the sample. First, we measure a reflection ($R_1$) and transmission ($T_1$) spectrum of the sample when there is no pump beam. We then measure the same spectra ($R_2$ and $T_2$) when a pump beam excites the sample. The third measurement of the transmission and reflection spectra is again performed without the pump beam ($R_3$, $T_3$) to observe any permanent change. This process is repeated at 5 different points on the sample and then averaged. The permanent transmission change plotted in Figure 1d of the paper is obtained by averaging ($T_3/T_1$)-1 for each power. The dynamics of the reflectivity are obtained using ΔR/R = ($R_2/R_1$) -1 and the equivalent for ΔT/T. These dynamics are shown in Figure S5 and are processed in the same way as the below-threshold data to calculate the dielectric function.

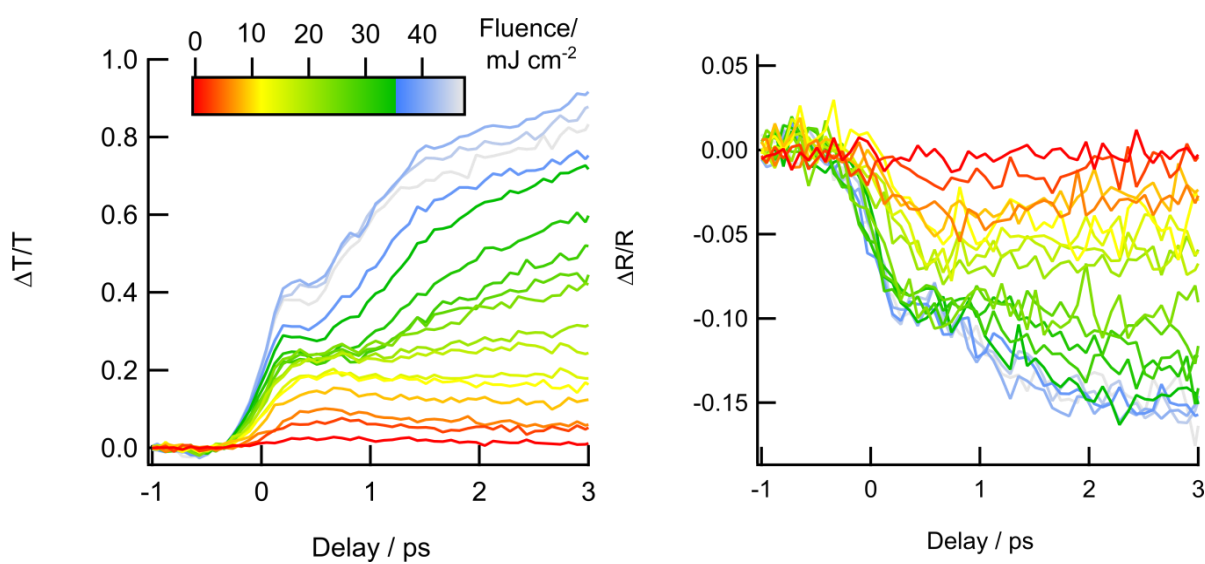

**Figure S5 | Measured ΔT/T and ΔR/R Time Traces Using the Single Shot Technique**

# BIBILOGRAPHY